\documentclass{article}

\usepackage{arxiv}

\usepackage[utf8]{inputenc} % allow utf-8 input
\usepackage[T1]{fontenc}    % use 8-bit T1 fonts
\usepackage{hyperref}       % hyperlinks
\usepackage{url}            % simple URL typesetting
\usepackage{booktabs}       % professional-quality tables
\usepackage{amsfonts}       % blackboard math symbols
\usepackage{nicefrac}       % compact symbols for 1/2, etc.
\usepackage{microtype}      % microtypography
\usepackage{lipsum}
\usepackage{graphicx}
\usepackage{caption}
\usepackage{multirow}
\usepackage{bm}
\usepackage{amsmath}
\usepackage{xurl}  % Allows automatic breaking of long URLs
\usepackage{hyperref}  % Ensures URLs are clickable
\graphicspath{ {./} }

\title{Reinforcement Learning-Based Controlled Switching Approach for Inrush Current Minimization in Power~Transformers}

\author{Jone Ugarte Valdivielso, Jose I. Aizpurua, Manex Barrenetxea and Brian G. Stewart, % <-this % stops a space
	\thanks{J. Ugarte \& M. Barrenetxea are with the Electronics \& Computing Science Department, Mondragon Unibertsitatea, Arrasate, Spain (e-mail: \{jugarte,
mbarrenetxeai\}@mondragon.edu); \\ J. I. Aizpurua is with the Department of Computer Science and Artificial Intelligence, University of the Basque Country (UPV/EHU) and Ikerbasque, Basque Foundation for Science, Bilbao, Spain (e-mail: joxe.aizpurua@ehu.eus); \\ B. G. Stewart is with the University of Strathclyde, Glasgow, UK (e-mail: brian.stewart.100@strath.ac.uk); \\
“This work has been submitted to the IEEE for possible publication. Copyright may be transferred without notice, after which this version may no longer be accessible.”}
}

\begin{document}

\maketitle

\begin{abstract} 
    Transformers are essential components for the reliable operation of power grids. The transformer core is constituted by a ferromagnetic material, and accordingly, depending on the magnetization state, the energization of the transformer can lead to high magnetizing inrush currents. Such high amplitudes shorten the life expectancy of a transformer and cause power quality issues in power grids. Various techniques have been proposed to minimize the inrush current; however, the application of Reinforcement Learning (RL) for this challenge has not been investigated. RL incorporates the ability to learn inrush minimization strategies adjusted to the dynamic transformer operation environment. This study proposes an inrush current minimization framework by combining controlled switching with RL. Depending on the opening angle of the circuit breaker and the remanent fluxes at disconnection, the proposed method learns the optimal closing instant of the circuit breaker. Two RL algorithms have been trained and tested through an equivalent duality-based model of a real 7.4 MVA power transformer. The evaluation of the RL algorithms is carried out with real measurement data and compared with real laboratory inrush currents. The results show that the inrush current is effectively minimized with the proposed framework.

\end{abstract}

% keywords can be removed

\section{Introduction}
\label{sec:Intro}

Transformers are key components in the reliable operation of the power system and the integration of renewable energy sources (RES). However, their operation can present challenges. During transformer energization, high magnetizing currents known as inrush currents can appear. These inrush currents, which can have an amplitude several times greater than the nominal current of the transformer, occur as a result of the saturation of the ferromagnetic core. The amplitude of the inrush current depends on the magnetic characteristics of the transformer core, the remanent flux at the transformer during its connection, and the connection instant of the circuit breaker (CB). Although inrush currents are not as severe as faults, they can cause mechanical stresses in the transformer, power quality problems in the power grid, and the tripping of nearby CBs~\cite{Blume1944}. Moreover, the inrush current transient contains harmonics that, especially in weak grids, can potentially lead to ferroresonance effects if they coincide with system resonance frequencies \cite{Sepasi2023}. This phenomenon can cause temporary overvoltages in the grid, which can increase the risk of equipment insulation failure and system instability~\cite{Toki2007}.   

The presence of weak grids in the electrical distribution system is increasing due to the decentralization of power generation. The growing integration of transformer-connected RES in new sites that are far from consumption challenges the power quality and grid stability. In this context, inrush current mitigation techniques are becoming increasingly important \cite{Alassi2022}. The most common inrush current minimization strategy is controlled switching, which involves controlled disconnection and reconnection of the CB between the transformer and the power system. The controlled switching strategy is based on closing the breaker at the optimal instant for the minimum inrush current. The execution of this strategy differs depending on the transformer characteristics, the remanent flux, and the number of independently controllable poles in the CB. 

According to the comparison in \cite{Cano2017}, the best-controlled switching performance is obtained when using independent pole-operated CBs. However, three-pole-operated CBs are more common in the power system due to the higher cost of independent-pole CBs. Controlled switching with three-pole-operated CBs is analyzed in \cite{Pachore2021}, with inrush current peak results below the rated transformer current. To obtain the lowest inrush current, the CB connection should be made at the instant where the difference between the (i) remanent flux and (ii) expected magnetizing flux is minimum. Remanent flux is normally obtained by integrating the primary side voltage at the transformer disconnection instant. Some studies suggest that a slow decay in the amplitude of the remanent flux may occur when the transformer is disconnected for long periods \cite{Chiesa2010}. However, there are no definitive conclusions on the evolution over time of the remanent flux~\cite{Cano2015}.

In order to develop an inrush current minimization technique, it is necessary to have a precise model of the system under study. The development of an accurate power transformer model capable of reproducing real inrush current transients entails various challenges. Firstly, special attention should be paid to the modelling of the ferromagnetic iron core of the transformer. Normally, for inrush current transients, the hysteresis curve of the transformer core is represented by the Jiles-Atherton (JA) model \cite{Jiles1984, Jiles1986}. This model consists of a partial differential equation that contains five physical parameters that must be estimated \cite{Ugarte24}. Lastly, the electrical model of the power transformer also plays an important role in transformer modeling. In this context, the duality-based transformer model has shown satisfactory results in different inrush current simulation studies \cite{Chiesa2010, Chiesa2010_2, Mork2007}. 
This transformer model merges the electrical and core representations into a single model. Moreover, it accurately represents the iron core of the transformer by individually considering the saturation characteristics of each leg and yoke of the transformer.  

Due to the growing complexity of the power grid, an emerging trend to tackle sequential decision-making problems is the implementation of Reinforcement Learning (RL) based solutions \cite{Sutton2015}. RL is a machine learning method that consists of an agent that can make decisions within an environment based on its past knowledge and the received feedback. Depending on the outcome of the decision, the agent receives a reward that is used to train a Neural Network (NN) \cite{Esmaeili2024}. NNs learn to adopt optimal actions that maximize the given reward by interacting with the environment. RL has been used to solve problems related to smart grids \cite{Chen2022}, such as frequency regulation \cite{Khooban2021, Chen2021}, voltage control \cite{Xu2020, Yang2020, Wang2020}, and energy management \cite{Silva2020, Bui2020, Qian2020}.  The ability to learn optimal actions based on the dynamics of the operation environment has led to the proliferation of RL solutions to solve issues related to power systems. However, to the best of the knowledge of the authors, RL has not been used to address power transformer inrush minimization strategies. Existing machine learning studies in the transformer and inrush current literature focus on discriminating inrush transients from fault currents \cite{Afrasiabi2022, Shu2022}, but inrush minimization studies have not been~addressed. 

In this research, RL and controlled switching are combined to minimize the inrush current in power transformers. Since the availability of real transformer data for detailed modelling is limited, this research work is focused on a 7.4 MVA 30/20 kV power transformer located in the medium voltage laboratory of Mondragon University \cite{Mondragon}. Through the RL method, the agent learns the best CB connection instant depending on the CB opening angle and estimated remanent fluxes. For this aim, the transformer under study is modelled by the duality-based approach and the JA core method. The proposed transformer model is validated by comparing real-case inrush current with simulation results for different current amplitudes. 
The proposed inrush minimization strategy is tested with two RL algorithms: Deep $Q$-Network (DQN) and Proximal Policy Optimization (PPO), including two variants of the DQN algorithm. The RL algorithms are trained with the remanent fluxes obtained by the validated duality-transformer model. The trained NN is evaluated with real-case remanent fluxes. These real remanent fluxes are obtained from classical non-controlled CB closing instants and serve as the base-case of this study. Lastly, the validation of the inrush current minimization technique is completed by comparing the inrush current results from the RL algorithm evaluations with the base-case inrush current measurements. 

The transformer environment model used in the RL algorithms and additional materials, including detailed explanations of the experimental setup, are available at \cite{Github2025}. These supplementary materials provide comprehensive support for the further exploration and implementation of different inrush minimization strategies in the same environment.

The remainder of the paper is organized as follows. Section \ref{sec:fundamentals_and_proposed_approach} introduces the basics behind transformer modelling and the RL method, along with the proposed approach for developing and evaluating the strategy. Then Section \ref{sec:case_study} presents the transformer model validation along with the training, evaluation, and validation of the proposed inrush current minimization strategy. In Section \ref{sec:Conclusions} the conclusions are presented.

\section{Fundamentals and Proposed Approach}
\label{sec:fundamentals_and_proposed_approach}

\subsection{Basics of Transformer Modelling}
\label{ssc:Transformer_modelling}

\begin{figure}[ht]
    \centering
    \includegraphics[width=0.45\linewidth]{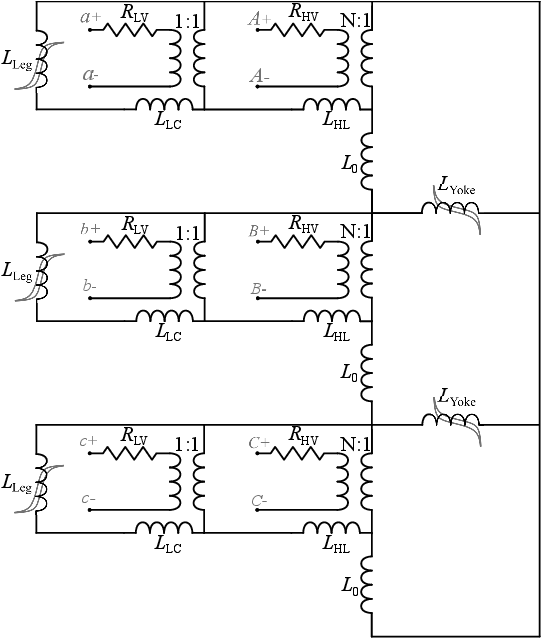}
    \caption{Duality-based transformer electrical circuit of a three-phase transformer \cite{Chiesa2010_2}.}
    \label{fig:Trafo_model}
\end{figure}

The electrical transformer model is based on an equivalent circuit derived from the duality transformation, which is based on the transformer magnetic circuit \cite{Martinez2005}. The model combines electrical parameters such as winding resistances and leakage inductances with an accurate representation of the transformer core. This makes the duality transformer model appropriate for electromagnetic transient assessments and inrush studies. Fig.~\ref{fig:Trafo_model} shows the duality-based electrical circuit of a three-phase, three-legged transformer with two windings.

\vspace{-0.5cm}
$R_{\text{HV}}$ and $R_{\text{LV}}$ are the resistances of the high- and low-voltage windings. $L_{\text{HL}}$ and $L_{\text{LC}}$ are the leakage inductance between the high- and low-voltage windings and between the low-voltage winding and the core. $L_0$ is the zero-sequence inductance. $L_{\text{leg}}$ and $L_{\text{yoke}}$ are the non-linear inductances that represent the magnetic characteristics of the core legs and core yokes with the JA model. The implementation of the JA model consists of five physical parameters that need to be estimated. A more detailed explanation of parameter estimation can be found in \cite{Ugarte24}. In this study, the core is made of an M120-27S material. The $\Delta$ and y connections, the phase shift, and the parasitic capacitance network are made externally connecting them to the transformer terminals identified with $a$, $b$ and $c$ letters (uppercase and lowercase). The main data of the transformer under study and the values of the parameters shown in Fig. \ref{fig:Trafo_model} are presented in Table \ref{tab:t_para_basic}. 

\begin{table}[ht]
\small
\caption{Main parameters of the transformer under study.}
\label{tab:t_para_basic}
\centering
\begin{tabular}{lccc}
\toprule
\textbf{Parameter} & \textbf{Symbol} & \textbf{Value} & \textbf{Unit} \\ \midrule
Apparent Power     & \textit{S}               & 7.4            & MVA           \\ 
Primary Voltage    & $V_{\text{p}}$    & 30             & kV            \\ 
Secondary Voltage  & $V_{\text{s}}$    & 20          & kV            \\ 
Primary Current    & $I_{\text{p}}$     & 142.4          & A             \\
Frequency          & \textit{f}               & 50             & Hz            \\ 
Primary Turns      & \textit{N}               & 824            & -             \\ 
High-Voltage Winding Resistance         &    $R_{\text{HV}}$    &   1.28      & $\Omega$         \\ 
Low-Voltage Winding Resistance         &    $R_{\text{LV}}$    &   0.26      & $\Omega$            \\ 
Low-Voltage to Core Leakage Inductance &$L_{\text{LC}}$    &  55.58  &    mH   \\ 
High- to Low-Voltage Leakage Inductance &$L_{\text{HL}}$  &  81.12  &    mH   \\ 
Zero-Sequence Inductance & $L_0$ & 25.37 & $\mu H$ \\
\bottomrule
\end{tabular}
\end{table}

\subsection{Basics of Reinforcement Learning}
\label{ssc:RL}

RL algorithms solve sequential decision-making problems that require a series of actions to obtain an optimal solution. The algorithm follows a Markov Decision Process (MDP) and as such is typically represented through the tuple $\langle \mathcal{S}, \mathcal{A}, \mathcal{R}, \mathcal{P}, \gamma \rangle$, where $\mathcal{S}$ represents the state of the environment, $\mathcal{A}$ represents the set of actions that RL can take,  and $\mathcal{P}$ represents the transition probability function. The reward function is typically represented as a function of the next state and the current action $\mathcal{R}(s_{t+1},a_t)$. The function generates the reward due to the action induced by the state-transition from $s_{t}$ to $s_{t+1}$. Finally, $\gamma \in [0,1]$ is the discount factor that determines the relevance of the rewards \cite{Sutton2015}.
 
\begin{figure}[ht]
    \centering
    \includegraphics[width=0.65\linewidth]{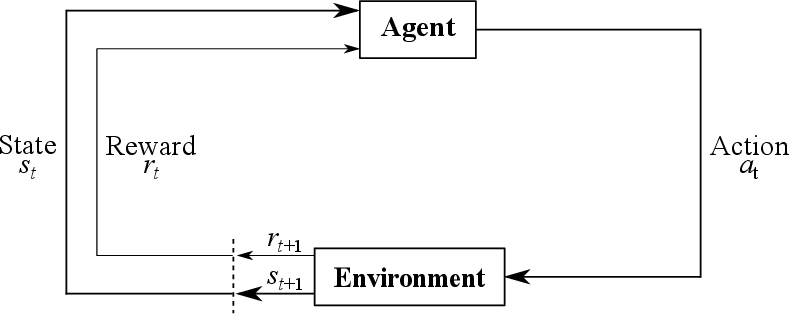}
        \caption{Agent-environment interaction in RL \cite{Sutton2015}.}
    \label{fig:RL_basecs}
\end{figure}

RL is based on the decision-making of an agent in a dynamic environment to maximize a cumulative reward. At each time step ($t$), the agent takes an action ($a_t$) depending on the state ($s_t$) and its past knowledge. This action is applied to the environment, providing the agent with a reward for the next state ($r_t$) based on the quality of the action and the transition to the next state ($s_{t+1}$). The RL agent-environment interaction process is shown in Fig. \ref{fig:RL_basecs}. 

In this research, the modelled environment is a power transformer together with laboratory cables, CBs, and the power grid. The state corresponds to the opening angle of the CB and the three remanent fluxes, and the action to take is the closing instant of the CB. Thus, the agent learns the best CB closing instant (action) depending on the opening angle and the remanent fluxes (state) of the last CB opening. The learning is carried out by training an NN to follow an optimal policy, \textit{i.e.} the function that the RL algorithm learns to maximize the cumulative reward. The RL algorithms selected in this study are DQN and PPO due to the extensive use of DQN and the improved training stability of PPO \cite{Chen2022}.

\textbf{Deep $Q$-Network}. Conventionally, for small state-action spaces, the training in DQN is carried out with the help of a $Q$-table. Each row in the $Q$-table corresponds to a state-action pair $Q$-value. At each time step, the $Q$-table is updated by training a $Q$-function. Therefore, depending on the selected state-action pair by the agent at each time step, the $Q$-function searches the $Q$-table for the corresponding $Q$-value. At the beginning of the training, the $Q$-table does not provide any information about the state-action selection. Depending on the study, the state space can be too large, and the use of a $Q$-table can become inefficient. To overcome this limitation, in the present research work, NNs are used as function approximations instead of $Q$-tables~\cite{Wilson2023}.
    
To improve the current policy, the agent explores the environment through actions. The action of the agent for each state is chosen according to a $\varepsilon$-greedy policy. The $\varepsilon$-greedy method gives a 1-$\varepsilon$ probability of taking the action with the highest value for a given state, which is known as the exploitation phase. In contrast, the agent has an $\varepsilon$ probability of selecting a random action and performing the investigation, also known as the exploration phase. Normally, the $\varepsilon_{\text{initial}}$ value is set to 1 and the $\varepsilon$ value decays with each training time step. The decaying function can be linear or exponential, and the decaying rate depends on the hyperparameter $exploration fraction$. The $\varepsilon_{\text{final}}$ value is another hyperparameter that can be tuned. The selection of an adequate $\varepsilon$ decay function is essential for a proper exploration/exploitation trade-off in the training \cite{Huang2019}.

The value of a state-action pair $Q\left( s_t, a_t \right)$ is updated in each decision step following \cite{Silva2020}:

    \begin{equation}\label{eq:Q_function}
    \begin{aligned}
        Q\left( s_t, a_t \right) &\leftarrow Q\left( s_t, a_t \right) 
        &+ \alpha \left(r_t + \gamma \max_a Q \left(s_{t+1}, a \right) 
        - Q \left( s_t, a_t \right) \right)
    \end{aligned}
    \end{equation}

\noindent where $s_t$ and $a_t$ are the state and action in time step $t$, respectively, $\alpha$ is the learning rate, which determines how fast the NN is trained, $r_t$ is the reward in time step $t$, $\gamma$ is a discounting factor and $max_aQ \left(s_{t+1},a \right)$ is the best state-action pair value of the next state $s_{t+1}$ and a given action~$a$.

\textbf{Proximal Policy Optimization}. PPO is an $actor-critic$ RL algorithm, in which two networks are trained simultaneously. The first network, the $actor$, is a policy that gives the probability of taking an action over a state. The second network, the $critic$, learns to approximate a value-based function similar to DQN, and assists the policy update of the $actor$ function by providing feedback on the quality of the action taken. This technique reduces the variance in training for the same starting point compared to other RL algorithms~\cite{Bick2021}.
    
Moreover, PPO has another mechanism to avoid instability in its training. This is based on adding a clipping to the probability policy update in each time step to avoid parameter updates that can have a high policy change. The clipping is defined as follows \cite{Fraija2024}:  
    
    \begin{equation}\label{eq:clip}
    \text{clip} \left(p_t(\bm{\theta}), 1 - \epsilon, 1 + \epsilon \right) =
    \begin{cases}
    (1 - \epsilon) & \text{if } p_t(\bm{\theta}) < (1 - \epsilon) \\
    (1 + \epsilon) & \text{if } p_t(\bm{\theta}) > (1 + \epsilon) \\
    p_t(\bm{\theta}) & \text{else}
    \end{cases}
    \end{equation} 

    \noindent where $\epsilon$ is the clipping range and $p_t(\bm{\theta})$ is the probability ratio between the old and new policies, with $\bm{\theta}$ being the trainable parameters of the NN biases and weights. The probability ratio is defined as follows:

    \begin{equation}\label{eq:pt}
    p_t \left(\bm{\theta} \right) =\frac{\pi_{\bm{\theta} \left(a_t | s_t \right)}}{\pi_{\bm{\theta_{\text{old}}}\left(a_t | s_t \right)}}
    \end{equation} 

    \noindent where $\pi_{\bm{\theta} \left(a_t | s_t \right)}$ is the probability of taking action $a_t$ in a state $s_t$ in the current policy and $\pi_{\bm{\theta_{\text{old}}} \left(a_t | s_t \right)}$ is the same in the old~policy. 

    The clipped objective function $L^{\text{CLIP}}$ is calculated by means of (\ref{eq:clip}) and (\ref{eq:pt}) and is defined as follows:
    
    \begin{equation}\label{eq:L_clip}
    L^{\text{CLIP}} = \hat{\mathbb{E}}_t \left[ \text{min} \left( p_t (\bm{\theta})\cdot A_t, \text{clip} \left(p_t(\bm{\theta}), 1 - \epsilon, 1 + \epsilon \right) \right)\cdot A_t \right]
    \end{equation} 

    \noindent where $\hat{\mathbb{E}}_t$ is the expectation over an episode and $A_t$ is the advantage estimate, which expresses how good or bad it is to take an action in a state compared to the average value of that~state.

    In contrast, the value-based function, the $critic$ network, is trained by minimizing the difference between the predicted state value $V_{\omega} \left(s_t \right)$ and its target state value $V_t^{\text{target}}$. The value-based function $L^{\text{V}}$ is defined as follows:

    \begin{equation}\label{eq:L_V}
    L^{\text{V}} = \hat{\mathbb{E}}_t \left[ \left( V_{\omega} \left(s_t \right) - V_t^{\text{target}} \right)^2 \right]
    \end{equation} 
    
    Moreover, exploration and exploitation are handled differently compared to DQN. Exploration is encouraged by adding an entropy term, $H$, to the overall objective function defined as follows:

    \begin{equation}\label{eq:L_all}
    L^{\text{CLIP+H+V}} = L^{\text{CLIP}}+ hH + vL^{\text{V}}
    \end{equation}     

    \noindent where $L^{\text{CLIP+H+V}}$ is the overall objective function and $h$ and $v$ are hyperparameters used as weighting factors for the entropy and the value-based functions.

\subsection{Proposed Approach}
\label{ssc:Proposed_model}

\begin{figure*}[t]
    \centering
    \includegraphics[width=1\linewidth]{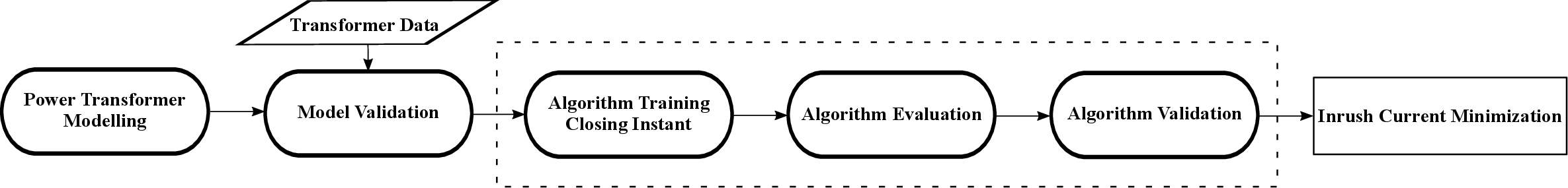}
    \caption{Overall block diagram of the proposed approach.}
    \label{fig:Flow_chart_general}
\end{figure*}

The proposed approach is based on the combination of controlled switching and RL to minimize the inrush current of power transformers. Fig. \ref{fig:Flow_chart_general} presents the steps to implement the proposed inrush minimization framework. The first step is to have an accurate representation of the system. Therefore, the study starts with the \texttt{Power Transformer Modelling} based on the duality transformation presented in Section \ref{ssc:Transformer_modelling}, which is implemented in Simulink. The \texttt{Model Validation} is performed by comparing the simulation inrush current results with real measured \texttt{Transformer Data} for different inrush current levels. 

\begin{figure}[ht]
    \centering
    \includegraphics[width=0.4\linewidth]{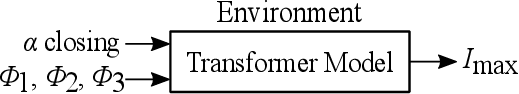}
    \caption{Diagram of the environment.}
    \label{fig:environment}
\end{figure}

The three stages covered by the dashed box in Fig.~\ref{fig:Flow_chart_general} are part of the RL inrush current minimization strategy framework. Once the transformer model is validated, the RL method is used to \texttt{train the NN for the optimal closing instant}. For the training phase, opening angles and remanent fluxes obtained from the validated simulation transformer model are used. These cases are generated by simulating the CB opening for 360$^{\circ}$ with 1$^{\circ}$ steps and obtaining the remanent fluxes for each step. Note that the opening and closing angles are defined with respect to the positive zero-crossing voltage of phase $a$. For each iteration of the training process, the remanent fluxes at that state ($\phi_1, \phi_2, \phi_3$), and the selected closing angle (action) by the RL algorithm are used to feed the transformer model (environment). The model responds with the maximum peak inrush current value ($I_{\text{max}}$) in~[pu] for that state-action pair, and with this data, the reward is calculated. The diagram in Fig. \ref{fig:environment} shows the inputs and output of the environment.

\vspace{0.5cm}

The objective of the RL algorithm is to maximize the cumulative reward, and therefore the inrush current minimization strategy should be able to minimize the peak at least below the rated current at 1 pu. The reward function is defined as follows:

\begin{equation}\label{eq:reward}
r =
\begin{cases} 
    -I_{\text{max}}, & \text{if } I_{\text{max}} > 1, \\
    (1-I_{\text{max}}), & \text{otherwise.}
\end{cases}
\end{equation}

\noindent where $r$ is the reward. 

Moreover, two algorithms, DQN and PPO, are tested to train the NNs. In DQN, two types of $\varepsilon$-greedy policies are tried to obtain the best strategy for a proper exploration/exploitation trade. The compared $\varepsilon$-greedy policies have linear and exponential decays.

The learning process of these RL algorithms is controlled by hyperparameters, which control the learning steps, the training speed, and the proper convergence of the algorithm. Therefore, setting the hyperparameters appropriately is an important step in the modelling process. In this study, hyperparameter selection is conducted by Bayesian optimization, which is implemented through the Optuna library \cite{Akiba2019}. The different hyperparameters used for the DQN and PPO algorithms, their tuning range, and their selected values are presented in TABLE~\ref{tab:hyperparameter}.

\begin{table*}[ht]
\centering
\caption{Hyperparameter ranges for DQN and PPO tuning.}
\label{tab:hyperparameter}
\small
\begin{tabular}{l |c c c| c c}
\toprule
\multirow{2}{*}{\textbf{Hyperparameter}} & \multicolumn{3}{c|}{\textbf{DQN}} & \multicolumn{2}{c}{\textbf{PPO}} \\
                                       & \textbf{Range} & \textbf{Linear Select} & \textbf{Exponential Select} & \textbf{Range} & \textbf{Select} \\
\midrule
Batch Size                           & [32, 256]      & 256            & 256                & [32, 256]      & 256            \\
Clipping Parameter ($\varepsilon_{\text{clip}}$) & -  & -              & -                 & [0.1, 0.4]     & 0.14           \\
Discount Factor ($\gamma$)            & 0.99           & 0.99           & 0.99              & 0.99           & 0.99           \\
Entropy Coefficient ($h$)             & -              & -              & -                 & $\left[ \text{10}^{\text{-6}}, \text{10}^{\text{-1}} \right]$ & 5.53 $\times \text{10}^{\text{-2}}$ \\
Exploration Initial Epsilon ($\varepsilon_{\text{initial}}$) & 1   & 1             & 1                 & -              & -              \\
Exploration Final Epsilon ($\varepsilon_{\text{final}}$)   & $\left[\text{10}^{\text{-5}}, \text{10}^{\text{-2}}\right]$ & 9.19 $\times \text{10}^{\text{-4}}$  & 8.67 $\times \text{10}^{\text{-4}}$ & -              & -              \\
Exploration Fraction                 & $\left[\text{10}^{-2}, 1\right]$ & 0.22          & 0.49               & -              & - \\
Learning Rate ($\alpha$)             & $\left[\text{10}^{-4}, \text{10}^{\text{-2}}\right]$ & 2.74 $\times \text{10}^{\text{-3}}$ & 1.15 $\times \text{10}^{\text{-3}}$ & $\left[\text{10}^{\text{-4}}, \text{10}^{\text{-2}}\right]$ & 2.42$ \times \text{10}^{\text{-3}}$ \\
Replay Buffer Size                    & $\left[\text{10}^{\text{3}}, \text{10}^{\text{5}}\right]$ & 1000         & 100000             & -              & -              \\
Value-Based Coefficient ($h$)         & -              & -              & -                 & [0.1, 0.9]     & 0.89 \\
\bottomrule
\end{tabular}
\end{table*}
\vspace{1cm}
The training stop criterion is set to 70,000 iterations. The selection of this value is justified from the search space of the problem, which consists of 360 opening angles with 360 closing angles, resulting in a total of 129,600 different cases. Consequently, the maximum iteration number is set to a value close to half of the total search space.

\begin{figure}[ht]
    \centering
    \includegraphics[width=0.5\linewidth]{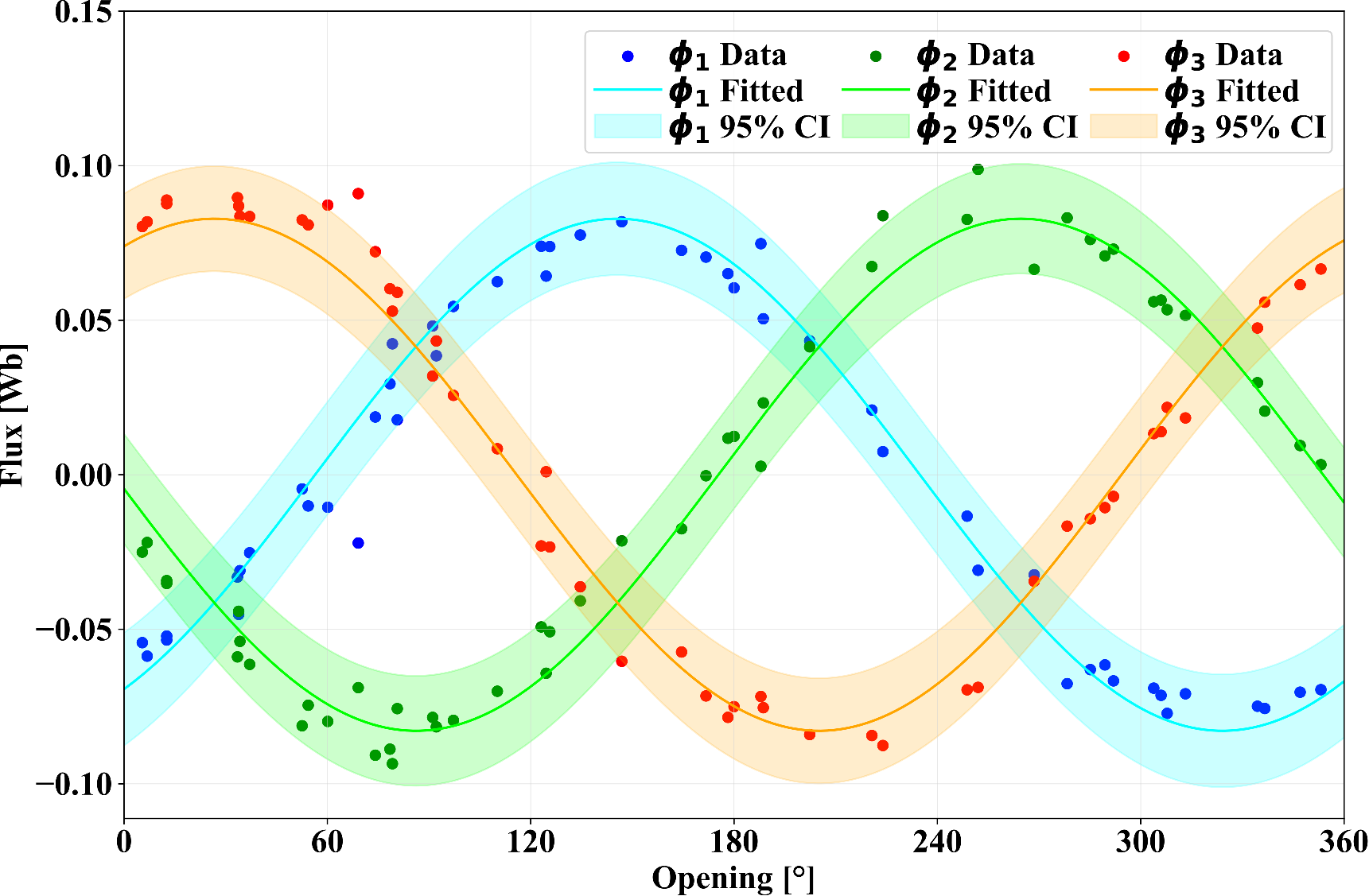}
    \caption{Real remanent flux data, fitted curve and 95$\%$ CI.}
    \label{fig:fitted_simu}
\end{figure}

The \texttt{evaluation} of the trained NN for each RL algorithm is carried out by first selecting and post-processing new data from the measurements. In this case, the new data are a set of 48 CB openings and three-phase remanent fluxes calculated by integrating the primary side voltage measurements of the transformer during the CB opening transients. There is no mechanism to control the opening instant of the CB, and therefore, openings can occur at any angle of the voltage waveform. The three-phase remanent fluxes, $\phi_1$, $\phi_2$ and $\phi_3$,  are obtained by integrating the primary side voltage of the transformer. Their values are presented in Fig. \ref{fig:fitted_simu} as data points with respect to their opening angle. These data are approximated by curve fitting sines, and their Confidence Interval (CI) of 95$\%$ is also shown. The equation that describes the evolution of the remanent fluxes concerning the opening angle is expressed as follows:
\begin{equation}\label{eq:flux}
\phi = A \sin{ \left( \omega t + \psi \right)} \pm \delta 
\end{equation} 
\noindent where $\phi$ is the remanent flux in [Wb] and $A$ is its amplitude in [Wb]. $\omega$ corresponds to 2$\pi$$f$, where $f$ is the 50 Hz frequency, in [rad/s], $t$ corresponds to the time in [s] and $\psi$ corresponds to the phase shift in [rad]. $\delta$ presents the tolerance between the data points and the fitted curve in [Wb]. The values of these parameters for the three remanent flux curves presented in Fig. \ref{fig:fitted_simu} are given in TABLE \ref{tab:fitting}.

\begin{table}[ht]
\centering
\caption{\textsc{Parameters for the fitted remanent flux curves.}}
\label{tab:fitting}
\begin{tabular}{lccccc}
\toprule
\textbf{Parameter} &\textbf{Symbol} & \textbf{$\phi_1$} & \textbf{$\phi_2$} & \textbf{$\phi_3$} & \textbf{Unit} \\ \midrule
{Amplitude} & A                  &  0.083  &  0.083     &  0.083 &  Wb     \\ 
{Phase Shift} & $\psi$          &   -0.9948        & 1.0995     &  3.1939     & rad \\
{Tolerance} & $\delta$            &    0.0093 &  0.0091   &  0.0087   & Wb                     \\ \bottomrule
\end{tabular}
\end{table}

\begin{figure}[ht]
    \centering
    \includegraphics[width=0.5\linewidth]{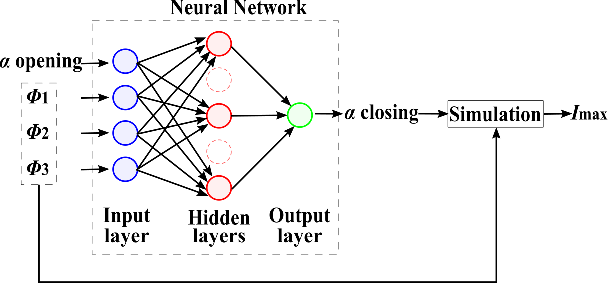}
    \caption{Visual representation of the \texttt{Evaluation} stage.}
    \label{fig:eval_exp}
\end{figure}

After obtaining the new data, the \texttt{Evaluation} is carried out as shown in Fig. \ref{fig:eval_exp}. The trained NN is fed with the opening angle and three-phase remanent fluxes (states). The NN gives an output (action) that is the closing angle of the CB. The remanent fluxes and the closing angle are used to run the transformer environment simulation model, and thus the inrush current peak for these settings is obtained. This process is repeated for the three variants of the RL algorithms (DQN linear, DQN exponential, and PPO), and their results are compared and analyzed. If the inrush current peak values are below 1 pu, training the RL algorithm is considered successful. 

The \texttt{RL algorithm validation} is performed by comparing the inrush current obtained with the trained RL algorithms and the laboratory tests performed with classical non-controlled CB closings. To validate the inrush current minimization strategy, 21 different inrush current measurements have been recorded. By comparing these real inrush currents with the results presented by the RL algorithms, the proposed approach is validated as an alternative \texttt{inrush current minimization} technique for power transformers.

\section{Numerical Results}
\label{sec:case_study}

The validation of the RL inrush current minimization strategy is performed by first validating the environment under study. The environment is the laboratory setup where the transformer under study is operated. The validation is carried out by comparing the inrush current values of the measurement with the simulation results of the transformer model. This environment is then used to train the agents of the selected RL algorithms. The RL strategy is analyzed in three parts. First, the training results are shown, followed by the evaluation of the trained NN with a new data set. Lastly, the proposed method is validated by comparing the results with the base-case scenario.

\subsection{Transformer Model Validation}

The transformer chosen to test and validate the RL inrush current minimization strategy is a 7.4 MVA 30/20 kV Dyn three-legged core transformer, which is energized from its primary side by the switching of a medium-voltage CB. The transformer model is based on duality transformation, as explained in Section \ref{ssc:Transformer_modelling}, and is implemented in Simulink to replicate inrush current transients. The nominal primary current for no-load conditions is 142.4 A rms. Taking into account this value, three different inrush current peak levels $I_{\text{max}}$ are chosen for the transformer model validation: 
\vspace{0.5cm}
\begin{itemize}
    \item Case 1: $I_{\text{max}}$ = 2.04 pu or 411 A 
    \item Case 2: $I_{\text{max}}$ = 0.85 pu or 172 A 
    \item Case 3: $I_{\text{max}}$ = 0.28 pu or 56 A 
\end{itemize}
\vspace{0.5cm}
Case 1 corresponds to the maximum expected inrush current according to the transformer manufacturer. Cases 2 and 3 are considered acceptable because their maximum value is below 1 pu. The comparison between the inrush currents obtained in simulation and in reality is shown in TABLE \ref{tab:trafo_validation}. 

\begin{table}[ht]
\centering
    \caption{\textsc{Difference between the real measured and simulated inrush currents.}}
\label{tab:trafo_validation}
\begin{tabular}{cccc}
\toprule
\textbf{Case} & \textbf{$I_{\text{difference}}$ [pu]} & \textbf{$I_{\text{difference}}$ [A]}  \\ \midrule
1             & 0.04           & 7.27                        \\ 
2             & 0.05                   & 11.04                              \\ 
3             & 0.11                   & 21.19                           \\ \bottomrule
\end{tabular}
\end{table}

\begin{figure}[!ht]
    \centering
        \includegraphics[width=0.5\columnwidth]{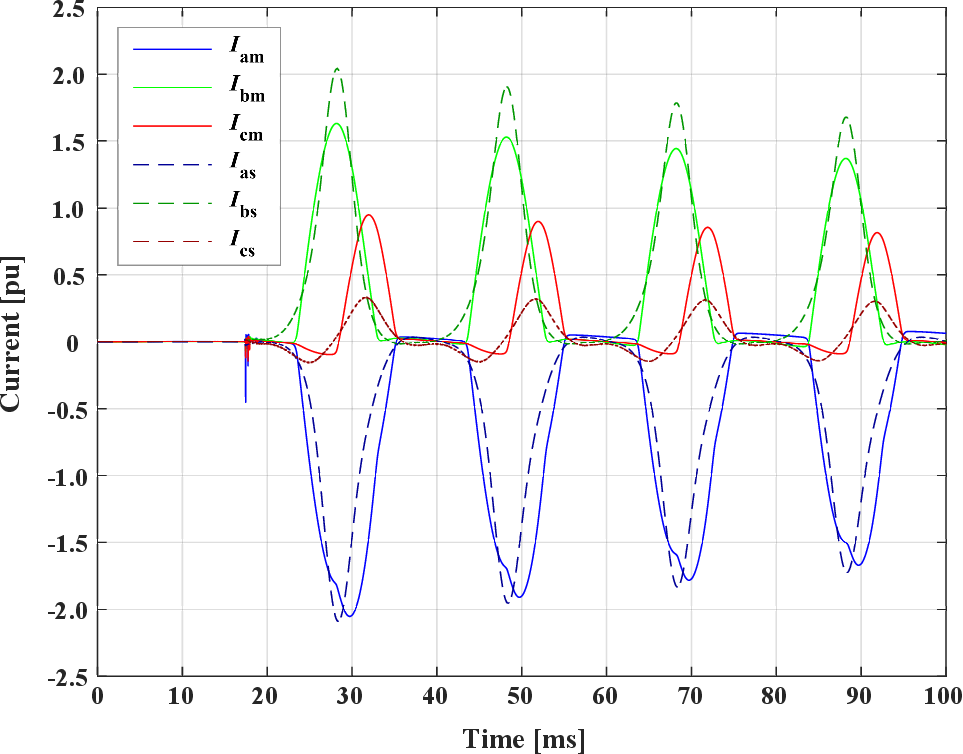}
    \caption{Comparison between measurement and simulink inrush current for Case 1.}
    \label{fig:I_validation1}
\end{figure}

\begin{figure}[!ht]
    \centering
        \includegraphics[width=0.5\columnwidth]{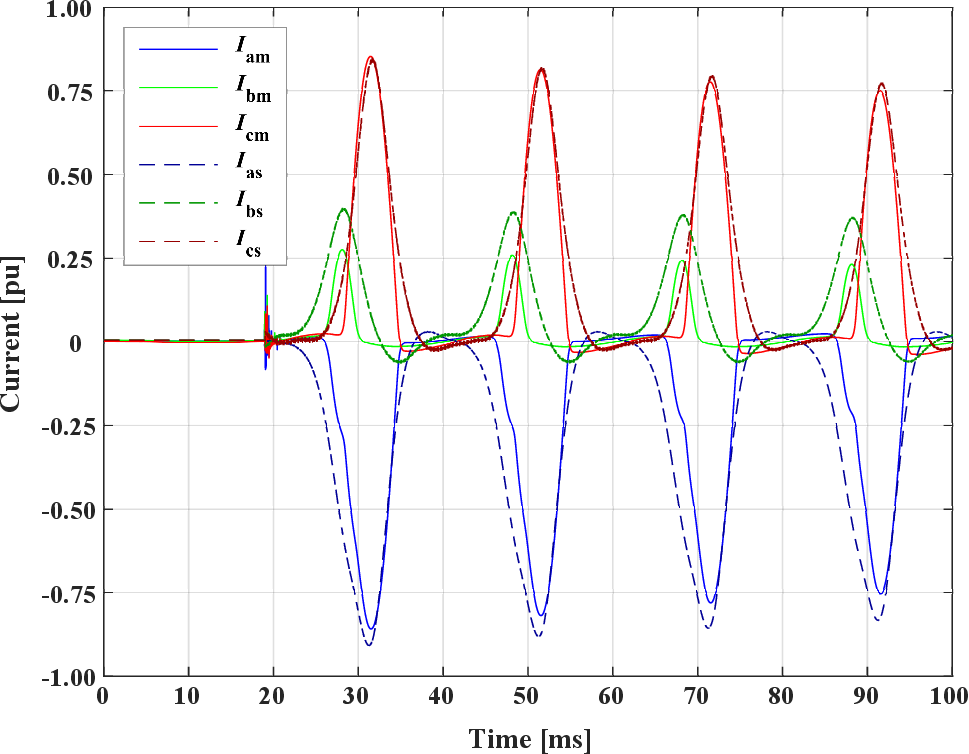}
    \caption{Comparison between measurement and simulink inrush current for Case 2.}
    \label{fig:I_validation2}
\end{figure}

\begin{figure}[!ht]
    \centering
        \includegraphics[width=0.5\columnwidth]{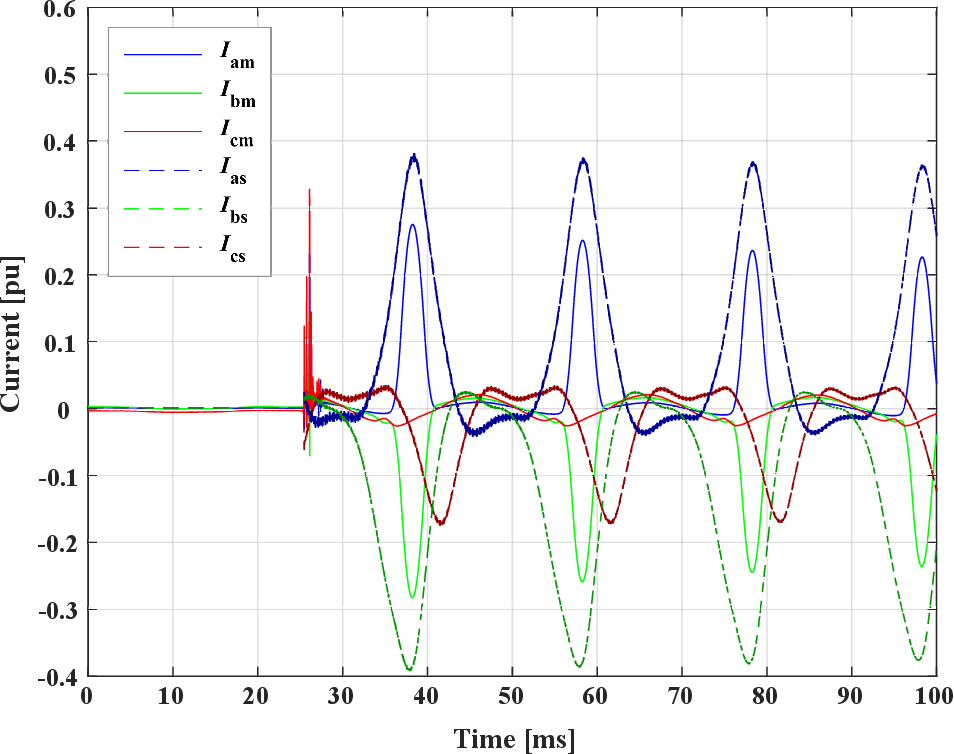}
    \caption{Comparison between measurement and simulink inrush current for Case 3.}
    \label{fig:I_validation3}
\end{figure}
\vspace{0.5cm}

The results show that the implemented transformer model is able to replicate the inrush current peak values obtained in the laboratory. Apart from that, the peak errors between simulation and measurement are 0.04 pu or 7.27 A for case 1, 0.05 pu or 11.04 A for case 2 and 0.11 pu or 21.19 A for case 3. The maximum error of 21.19 A occurs when the inrush current peak is minimum. This only means a 0.11 pu difference with respect to the real measurement. As the objective of this study is to train an RL algorithm to minimize the inrush current below 1 pu, this error is assumed to be acceptable. 

In addition to the inrush current peak, Figs. \ref{fig:I_validation1}, \ref{fig:I_validation2}, and \ref{fig:I_validation3} show the comparison between the inrush current waveforms obtained from the measurements and from the duality-based transformer model for cases 1, 2, and 3, respectively. The results demonstrate that the real and simulated inrush current waveforms have the same pattern. This indicates that the implemented duality-based transformer model can replicate the inrush current transients observed in the transformer under~study.

\subsection{Inrush Current Minimization Strategy Results}

\subsubsection{Training}

\begin{figure}[h!]
    \centering
    \includegraphics[width=0.6\linewidth]{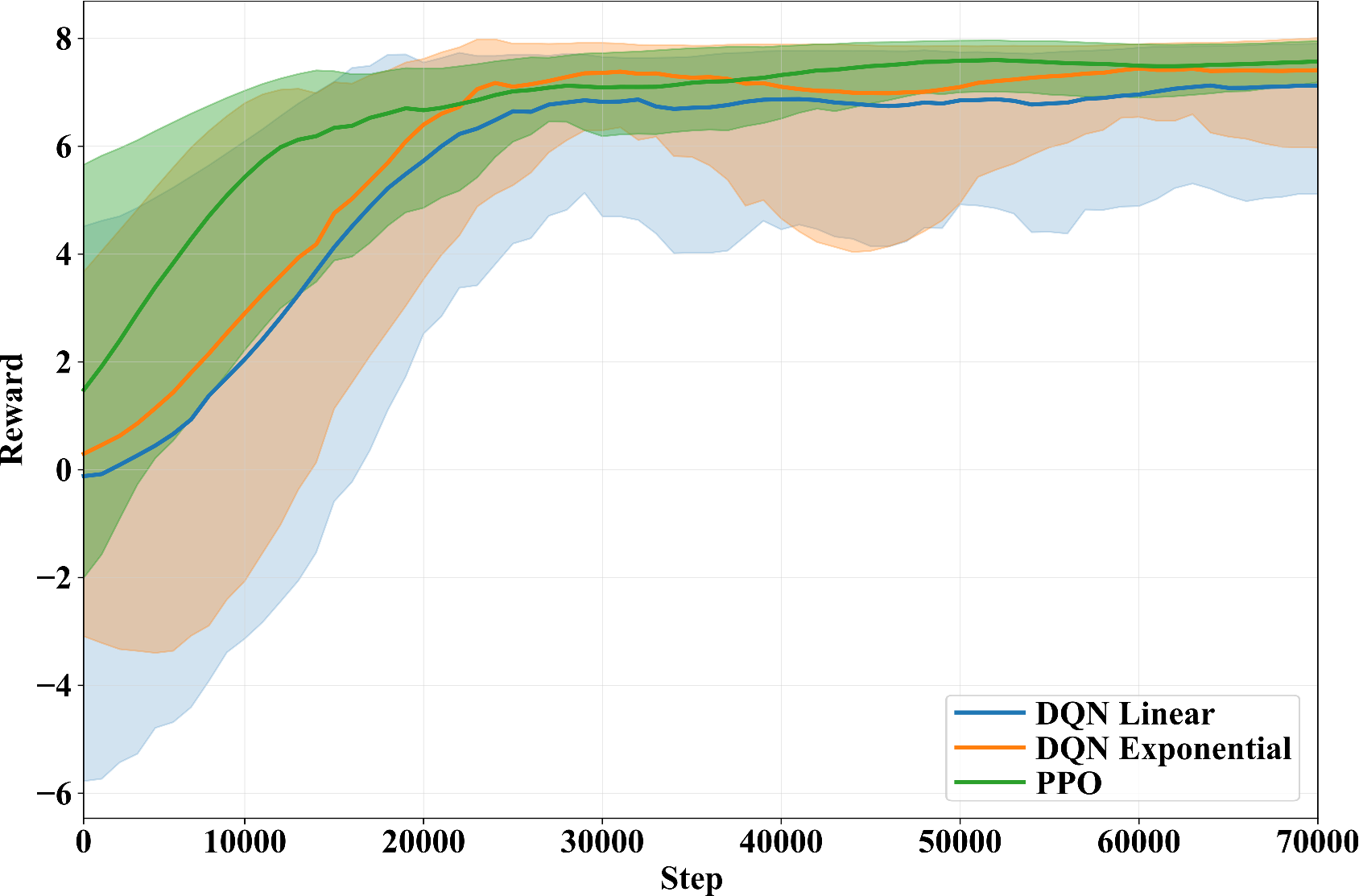}
    \caption{Average episode reward for each RL algorithm during training, shaded areas indicate 95$\%$ CI.}
    \label{fig:training_reward}
\end{figure}

The training process begins after the hyperparameters have been selected. Fig. \ref{fig:training_reward} shows the evolution of the training in each iteration of the average episode reward.  The episode reward is the cumulative sum of the rewards in an episode, and the rewards are calculated as described in (\ref{eq:reward}).  As shown in Fig. \ref{fig:training_reward}, PPO has the highest average episode reward and the lowest variance at the end of training. DQN with $\varepsilon$ exponential decay shows only a 2.5$\%$ decrease in the average episode reward, however, the variance at the end of the training is higher than for PPO. DQN with $\varepsilon$ linear decay shows the worst results with an average episode reward decrease of 5.5$\%$ compared to PPO and the highest variance at the end of training. Apart from the end of the training values, PPO also has the highest average episode mean and lowest variance at the beginning of the training process, while the two variants of DQN show a much higher variance in the first stages of the training. Although at some points the training of DQN with $\varepsilon$ exponential decay shows a higher average mean than PPO, the latter maintains a more stable evolution throughout the training. The instability of both DQN algorithms is evident from the high variance throughout its learning process. 

\vspace{0.5cm}

\subsubsection{Evaluation}

\begin{figure}[h!] % or [b] for bottom
    \centering
    \includegraphics[width=0.6\linewidth]{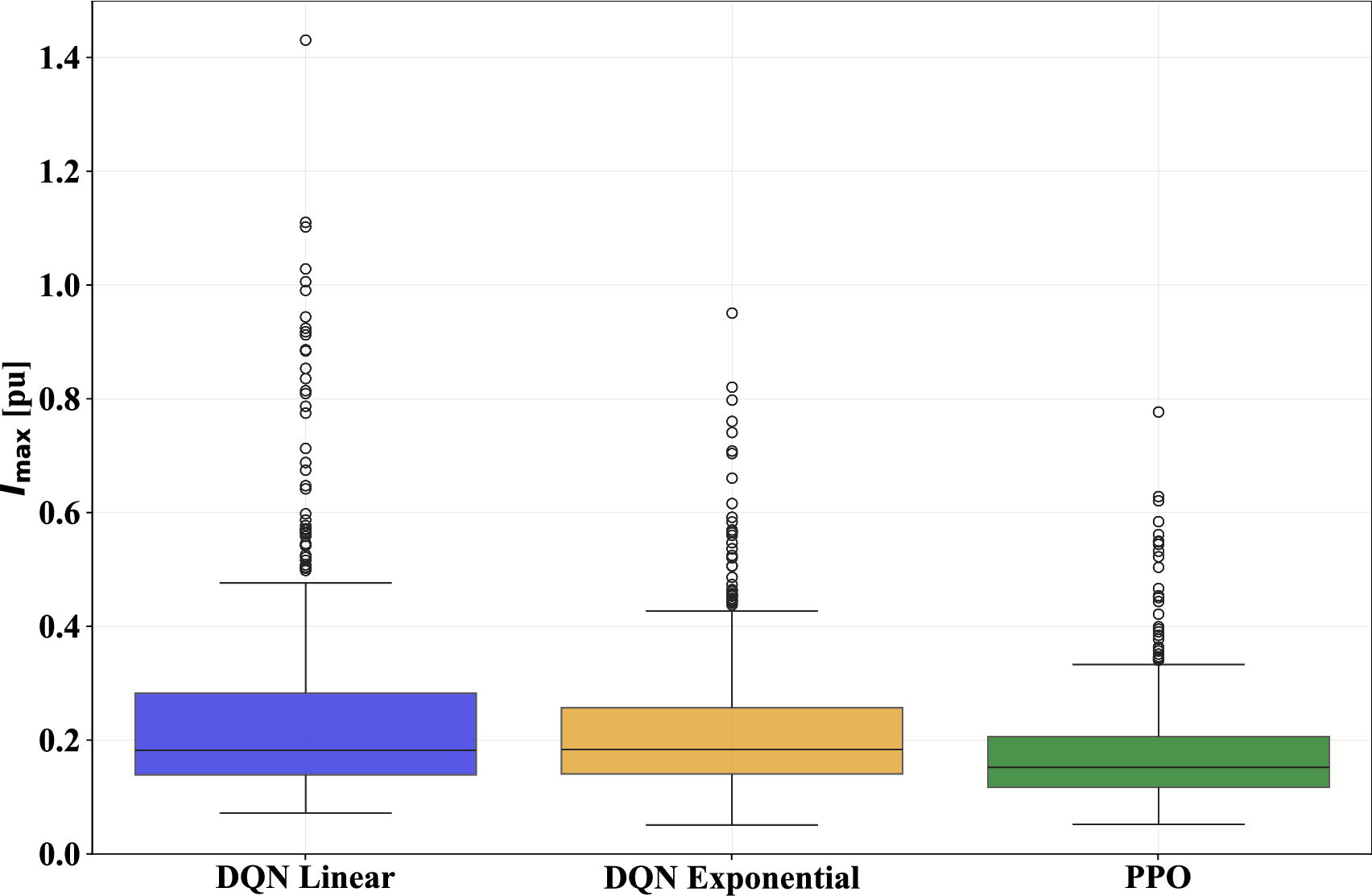} % \linewidth works the same here
    \caption{Box plot results of the evaluation for each RL algorithm.}
    \label{fig:boxplot}
\end{figure}

The inrush current peak results obtained in the evaluation stage for the different RL algorithms are visualized in Fig. \ref{fig:boxplot}, in the form of box plots. In the box plot, the DQN algorithm with $\varepsilon$ linear decay has outliers that exceed the acceptable limit of the inrush current, with a value greater than 1 pu. However, the overall mean value of the evaluation has an acceptable value of 0.25 pu. The evaluations of the other two algorithms show a lower mean of 0.22 pu and 0.18 pu for DQN with $\varepsilon$ exponential decay and PPO, respectively. The most significant difference between DQN with $\varepsilon$ exponential decay and PPO is their maximum value, which is higher in the first algorithm. However, both of these maximum inrush current peaks are within the 1 pu limit range and, hence, are considered acceptable. The most significant values of the box plot in Fig. \ref{fig:boxplot} are presented in detail in TABLE \ref{tab:boxplot}. The Q1 and Q3 parameters are the first and third quartiles and indicate where the 25$\%$ and 75$\%$ of the data are distributed. 

\begin{table}[h!]
\centering
\caption{\textsc{Inrush current values for the box plot in Fig. \ref{fig:boxplot}.}}
\begin{tabular}{ccccc}
\toprule
\textbf{Parameter}            & \textbf{DQN Linear} & \textbf{DQN Exponential} & \textbf{PPO} & \textbf{Units} \\ \midrule
Mean                       & 0.25                & 0.22                   & 0.18         & pu             \\ 
Median                       & 0.18                & 0.18                   & 0.15         & pu             \\ 
Q1                         & 0.14                & 0.14                    & 0.12         & pu             \\ 
Q3                         & 0.28                & 0.26                   & 0.21         & pu             \\ 
Minimum                    & 0.07                & 0.05                    & 0.05         & pu             \\ 
Maximum                    & 1.43                & 0.95                    & 0.78         & pu             \\ \bottomrule
\end{tabular}
\label{tab:boxplot}
\end{table}
\vspace{0.5cm}

The evaluation of the trained algorithms with new data shows that the training is satisfactory for DQN with $\varepsilon$ exponential decay and PPO. However, considering that the training of the RL algorithms is carried out to minimize the inrush current below 1 pu, and the DQN with $\varepsilon$ linear decay surpasses this limit, it is discarded for further analyses. 

\subsubsection{Validation}

\begin{figure}[ht] 
    \centering
    \includegraphics[width=0.6\linewidth]{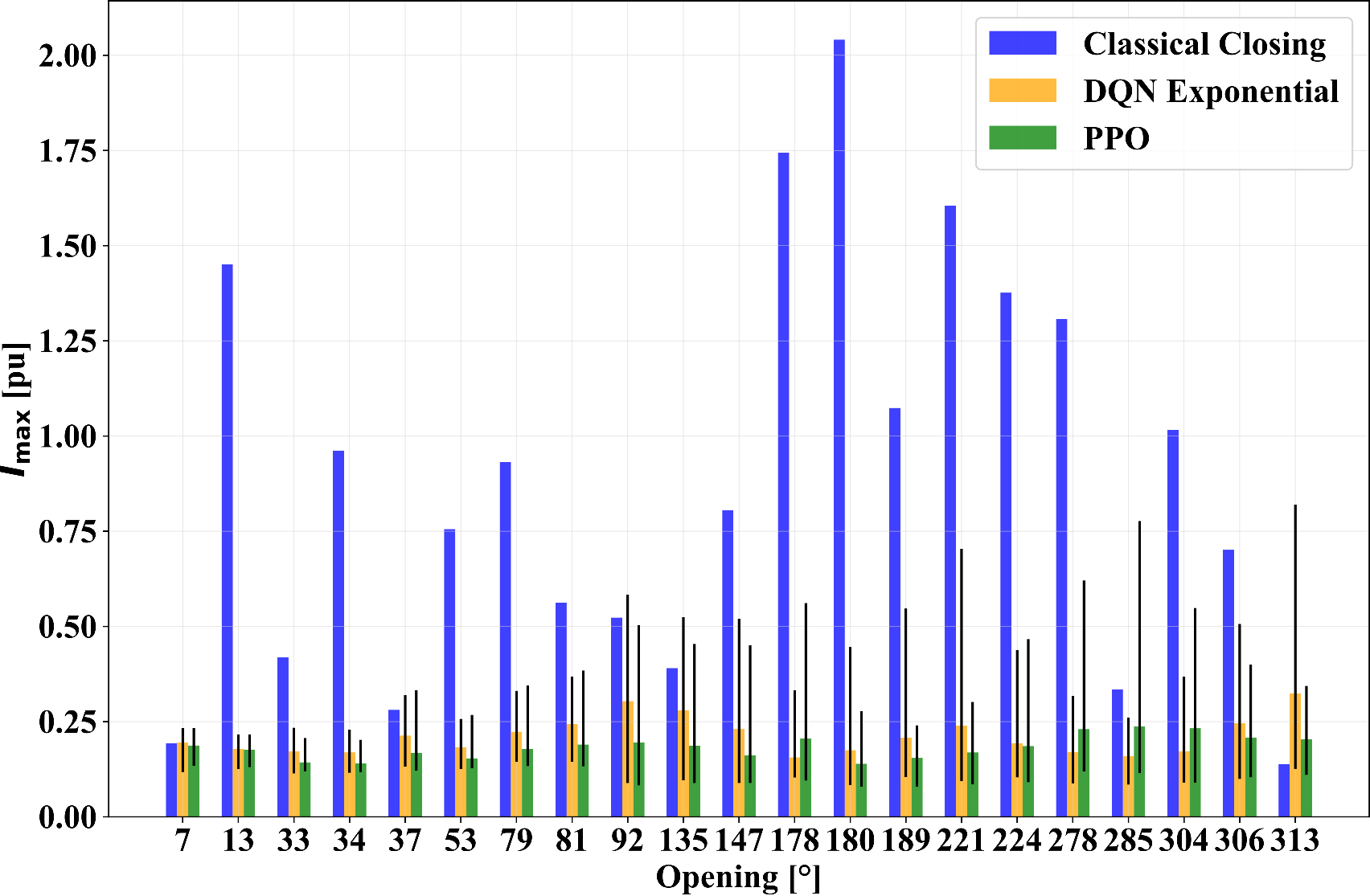} 
    \caption{Comparison between the inrush current peak from classical non-controlled closing and simulation, the lines indicate the minimum and maximum for each case.}
    \label{fig:Imax_validation}
\end{figure}

The inrush current validation results are shown along with their corresponding opening angles in Fig.~\ref{fig:Imax_validation}. The bar results indicate the mean, and the lines indicate the minimum and maximum inrush current values. These results are presented in more detail in TABLE~\ref{tab:imax_comparison}. The results show that with a classical non-controlled CB closing, the inrush current at the transformer connection exceeds 1 pu in 38$\%$ of the cases. When the proposed strategies are used, the maximum inrush current is always below 0.82 pu for DQN with $\varepsilon$ exponential decay and 0.78 pu for PPO. Furthermore, the maximum mean inrush current is 0.32 pu for DQN with $\varepsilon$ exponential decay and 0.24 pu for PPO. In 90$\%$ of the average cases, the peak inrush current obtained with the inrush current minimization strategy is lower than the real measured value, which shows the effectiveness of the proposed minimization~technique.

In addition to the individual values, the average inrush current peak is also calculated for the different cases. In the case of classical non-controlled closing, the average inrush current mean is 0.89 pu. DQN with $\varepsilon$ exponential decay and PPO indicate an average value of 0.21 pu and 0.18 pu, respectively. This results in a reduction of the average inrush current peak of 76$\%$ for DQN with $\varepsilon$ exponential decay and 80$\%$ for PPO compared to classical non-controlled closing. 

\begin{table}[ht]
\centering
\caption{\textsc{Mean, minimum and maximum of the inrush current results for each opening angle.}}
\begin{tabular}{c|c|ccc| cccc}
\toprule
\multirow{2}{*}{\textbf{Opening {[}$^\circ${]}}} & \multirow{2}{*}{\textbf{\begin{tabular}[c]{@{}c@{}}Classical\\ Closing {[}pu{]}\end{tabular}}} & \multicolumn{3}{c|}{\textbf{\begin{tabular}[c]{@{}c@{}}DQN\\ Exponential [pu]\end{tabular}}} & \multicolumn{3}{c}{\textbf{PPO [pu]}}             \\
                                         &                                                                                                & \textbf{Mean} & \textbf{Minimum} & \textbf{Maximum} & \textbf{Mean} & \textbf{Minimum} & \textbf{Maximum} \\
\midrule
7   & 0.19  & 0.19  & 0.12 & 0.23  & 0.19  & 0.13 & 0.23 \\ 
13  & 1.45  & 0.18  & 0.13 & 0.22  & 0.18  & 0.13 & 0.22 \\ 
33  & 0.42  & 0.17  & 0.11 & 0.23  & 0.14  & 0.12 & 0.21 \\  
34  & 0.96  & 0.17  & 0.12 & 0.23  & 0.14  & 0.12 & 0.20 \\  
37  & 0.28  & 0.21  & 0.13 & 0.32  & 0.17  & 0.12 & 0.33 \\  
53  & 0.76  & 0.18  & 0.13 & 0.26  & 0.15  & 0.13 & 0.27 \\  
79  & 0.93  & 0.22  & 0.15 & 0.33  & 0.18  & 0.13 & 0.35 \\  
81  & 0.56  & 0.24  & 0.14 & 0.37  & 0.19  & 0.13 & 0.38 \\  
92  & 0.52  & 0.30  & 0.09 & 0.58  & 0.20  & 0.08 & 0.50 \\  
135 & 0.39  & 0.28  & 0.10 & 0.52  & 0.19  & 0.09 & 0.45 \\  
147 & 0.81  & 0.23  & 0.09 & 0.52  & 0.16  & 0.09 & 0.45 \\  
178 & 1.74  & 0.16  & 0.10 & 0.33  & 0.21  & 0.10 & 0.56 \\  
180 & 2.04  & 0.17  & 0.08 & 0.45  & 0.14  & 0.08 & 0.28 \\  
189 & 1.07  & 0.21  & 0.10 & 0.55  & 0.15  & 0.08 & 0.24 \\  
221 & 1.61  & 0.24  & 0.09 & 0.70  & 0.17  & 0.09 & 0.30 \\  
224 & 1.38  & 0.19  & 0.10 & 0.44  & 0.19  & 0.09 & 0.47 \\  
278 & 1.31  & 0.17  & 0.09 & 0.32  & 0.23  & 0.12 & 0.62 \\  
285 & 0.33  & 0.16  & 0.08 & 0.26  & 0.24  & 0.12 & 0.78 \\  
304 & 1.02  & 0.17  & 0.09 & 0.37  & 0.23  & 0.09 & 0.55 \\  
306 & 0.70  & 0.25  & 0.10 & 0.51  & 0.21  & 0.10 & 0.40 \\  
313 & 0.14  & 0.32  & 0.13 & 0.82  & 0.20  & 0.11 & 0.34 \\  
\midrule  
\textbf{Mean} & 0.89 & 0.21  & 0.11  & 0.41  & 0.18  & 0.11  & 0.39 \\  
\bottomrule
\end{tabular}
\label{tab:imax_comparison}
\end{table}

\section{Conclusions}
\label{sec:Conclusions}

This study presents an inrush current minimization strategy for power transformers developed by combining RL and controlled switching. Two RL algorithms are trained with the proposed approach, DQN and PPO, including two variants of the DQN algorithm; $\varepsilon$ linear and exponential decays. The training of the algorithms is carried out using a duality-based transformer model. The objective of training is to learn the optimal CB closing angle based on the CB opening angle and the remanent fluxes at the transformer disconnection. The learning capacity of the proposed approach is evaluated by testing each RL algorithm with new data. The obtained inrush current results are compared with the base-case scenario, which is based on a widely adopted approach in the industry: classical non-controlled CB closing. The results show that the DQN with $\varepsilon$ exponential decay and PPO algorithms can minimize the inrush current below 1 pu. Furthermore, PPO is suggested to be the most stable and accurate inrush minimization solution due to the best results obtained with respect to the tested DQN~variants.

The proposed approach offers an alternative to classical inrush current minimization techniques. Moreover, the combination of transformer modelling and controlled switching with Reinforcement Learning opens the way to develop new transformer control and optimization strategies. The results obtained in this study show that, compared to the classical non-controlled CB closing technique, the implementation of the proposed framework reduces the average inrush current peak by 76$\%$ for the case of DQN with $\varepsilon$ exponential decay and 80$\%$ for PPO. Therefore, incorporating this approach into a real power transformer may help reduce transformer degradation and improve the stability of the power system. In addition, these features ease the integration of power transformer-connected RES into weak distribution grids, avoiding the voltage stability challenges inherited from high inrush currents.

The implementation of the proposed minimization technique for electrical power system-level studies would pose relevant challenges. For instance, although the transformer model presented in this study accurately represents the inrush current transients, the computation time is high. Therefore, to analyze inrush currents and other transients in system-level studies, a computationally efficient and versatile transformer model is necessary. This challenge can be addressed in the future with the development of a computationally efficient transformer surrogate model. 

\section{Acknowledgments}
This research was funded by the Spanish State Research Agency (grant No. CPP2021-008580) and Department of Education of the Basque Government, Research Group Program (grant No. IT1634-22 and IT1504-22). In addition, J. I. Aizpurua is funded by the Ramón y Cajal Fellowship, Spanish State Research Agency (grant number RYC2022-037300-I), co-funded by MCIU/AEI/10.13039/501100011033 and FSE+. Part of this work was undertaken by J. Ugarte Valdivielso at the University of Strathclyde, Scotland, UK, through an ERASMUS+ Student Mobility Award.

\end{document}